\documentclass[sigconf]{acmart}

\AtBeginDocument{%
  }

\acmConference[ACM RecSys '24]{Make sure to enter the correct
  conference title from your rights confirmation email}{October 14--18,
  2024}{Bari, Italy}
\usepackage{amsmath}
\usepackage{enumitem}
\usepackage{subcaption} 
\usepackage{multirow}
\usepackage{enumitem}

\begin{document}
\raggedbottom
\title[Optimizing Embedding Spaces for Culturally Balanced Recommendations]{Advancing Cultural Inclusivity: Optimizing Embedding Spaces for Balanced Music Recommendations}

\author{Armin Moradi}
\affiliation{%
  \institution{Mila Quebec AI Institute, Université de Montréal}
  \city{Montréal}
  \country{Canada}}
\email{armin.moradi@mila.quebec}

\author{Nicola Neophytou}
\affiliation{%
  \institution{Mila Quebec AI Institute}
  \city{Montréal}
  \country{Canada}}
\email{nicola.neophytou@mila.quebec}

\author{Golnoosh Farnadi}
\affiliation{%
  \institution{Mila Quebec AI Institute, Université de Montréal, McGill University}
  \city{Montréal}
  \country{Canada}}
\email{farnadig@mila.quebec}

\renewcommand{\shortauthors}{Moradi et al.}

\begin{abstract}

Popularity bias in music recommendation systems - where artists and tracks with the highest listen counts are recommended more often - can also propagate biases along demographic and cultural axes. In this work, we identify these biases in recommendations for artists from underrepresented cultural groups in prototype-based matrix factorization methods. Unlike traditional matrix factorization methods, prototype-based approaches are interpretable. This allows us to directly link the observed bias in recommendations for minority artists (the effect) to specific properties of the embedding space (the cause). We mitigate popularity bias in music recommendation through capturing both users’ and songs’ cultural nuances in the embedding space. To address these challenges while maintaining recommendation quality, we propose two novel enhancements to the embedding space: i) we propose an approach to filter-out the irrelevant prototypes used to represent each user and item to improve generalizability, and ii) we introduce regularization techniques to reinforce a more uniform distribution of prototypes within the embedding space. Our results demonstrate significant improvements in reducing popularity bias and enhancing demographic and cultural fairness in music recommendations while achieving competitive -if not better- overall performance.

\end{abstract}

\ccsdesc[500]{Information systems~Recommender systems}
\ccsdesc[300]{Information systems~Information retrieval}
\ccsdesc[100]{Social and professional topics~Computing / technology policy}
\keywords{music recommendation, popularity bias,  prototype-based matrix factorization, demographic fairness, cultural fairness}

\received{20 May 2024}
\maketitle

\section{Introduction}

Online music recommender systems have revolutionized the way in which we listen to and discover music, offering a personalized journey through a vast and diverse soundscape. However, these systems face a critical challenge: ensuring fair and unbiased recommendations across a global audience with a multitude of cultural preferences \cite{deldjoo2024fairness, bauer2018importance, lee2023understanding}.

This paper investigates the potential for cultural bias in music recommender systems, due to the influence of certain dominant cultures among artists within the data \cite{lesota2022traces}.  
This kind of bias is adjacent to demographic bias, but refers not just to systemic discrimination, but to how recommended content can become culturally homogeneous, particularly with the over-influence of Western styles. Such biases can limit users' exposure to a wider range of music, hindering their discovery of diverse cultural expressions \cite{bello2021cultural}.

While existing research has explored various strategies to mitigate popularity and demographic biases in music recommendation \cite{wei2021model, kamehkhosh2017user}, such as data pre-processing \cite{boratto2021interplay} and post-processing adjustments \cite{abdollahpouri2019managing, mansoury2021graph}, we examine an in-processing fairness approach \cite{wei2021model, zheng2021disentangling, rhee2022countering} to analyze the model's use of cultural nuances within the recommendation model itself. We leverage previous work on prototype-based recommender systems (ProtoMF) \cite{melchiorre2022protomf}, which transforms latent representations of users and items into new embedding spaces, defined by prototypes which act as anchor points. Our approach explicitly focuses on ensuring the model can learn from minority cultures, despite being less prevalent in the data. Further, since we deal directly with the embedding space, our approach has the additional benefit of transparency when ensuring equitable treatment of user subgroups. 
This work therefore also emphasizes the importance of explainable recommender systems, allowing users to understand the reasoning behind output recommendations and fostering trust \cite{afchar2022explainability}. We make the following key contributions:
\begin{enumerate}[leftmargin=*]
    \item We analyze the embedding space of the Prototype-based Matrix Factorization (ProtoMF) method to uncover the model's disparate treatment of cultures among items and subsequent popularity bias. 
    \item We then propose an extension to ProtoMF, \textbf{\textit{Prototype-$k$-Filtering}}, that uses only the $k$ closest prototypes to users and items in the embedding space to generate their representations.
    \item We propose a second extension to ProtoMF, \textbf{\textit{Prototype-Distributing Regularizer}}, which regularizes the loss function to evenly distribute prototypes throughout the embedding space. 
\end{enumerate}

With these contributions, we investigate the cultural information captured by representations of prototype-based recommendations, and how this can be augmented to enhance rather than deteriorate cultural fairness. Our extensive experimental results on two real-world datasets indicate that cultural bias in prototype-based recommender systems can be mitigated without any additional cost to utility. In this work, therefore, we aim to contribute to the development of more inclusive and equitable digital environments in the recommendation setting - ones which empower the work of creators from minority backgrounds and deliver curated and culture-aware diversity to consumers.   \textit{The dataset and code will be made publicly available upon publication of this work.}


\section{Related Work}
Mitigating popularity and demographic biases have both emerged as critical areas of research in recommender systems. 
In-processing fairness methods aim to integrate fairness considerations directly into the learning process of the recommendation system. These methods offer advantages in terms of transparency and control over the model's behavior \cite{burke2018balanced, patro2020fairrec}. One promising approach is fair matrix factorization, which incorporates fairness constraints into the objective function during model training \cite{rhee2022countering, wei2021model, zheng2021disentangling, rhee2022countering}. However, such approaches do not consider adapting properties of the embedding space indirectly. Our approach, which enhances prototype-based Matrix Factorization, introduces constraints to the learning process that are independent of the original loss function. This allows us to mitigate bias without sacrificing recommendation quality.

Incorporating prototypes into the learning process has emerged as a recent concept, primarily to address cold-start issues; \citet{sankar2021protocf} leverage prototypes for recommending long-tail items with sparse interactions as a few-shot learning problem with collaborative filtering (ProtoCF). A similar but alternative approach is identifying representatives for the cold-start problem. \citet{shi2017local} select local representative items from user groups using an interview-based model, while \citet{aleksandrova2017identifying} propose a method to  automatically interpret representative users. Anchor-based collaborative filtering (ACF), from \citet{barkan2021anchor}, shares this concept, requiring items to select single representative anchors. These approaches capture user/item characteristics with representative elements for recommendations with limited data.

While these prototype-based methods prove to be competitive due to capturing ``genre''-adjacent information, none have yet been considered for addressing cultural fairness issues. As such, we address these gaps in the existing work; we propose augmenting a prototype-based matrix factorization approach, ProtoMF \cite{melchiorre2022protomf}, and adapting properties of prototypes in the embedding space to promote cultural diversity among items and reduce popularity bias. Our approach is distinguished by its adaptations to the embedding space, which do not directly affect the loss function and therefore maintain competitive performance. While ProtoMF, without offering solutions, revealed gender bias among users, we extend this work by demonstrating cultural bias among items and propose solutions to mitigate both cultural and popularity biases. 


\section{Preliminaries}
\subsection{Prototype-based Matrix Factorization}
Music recommender systems typically rely on matrix factorization techniques to model user-item interactions. In this case, a user-item matrix with $N$ users and $M$ items ($R \in \mathbb{R}^{N \times M}$) is decomposed into user factor matrix ($U \in \mathbb{R}^{N \times d}$) and an item factor matrix ($T \in \mathbb{R}^{d \times M}$) where $d$ is the embedding space dimension \cite{dziugaite2015neural}. These factor matrices represent latent factors that capture user preferences and item characteristics separately. The affinity score can be calculated between user and item representations in order to capture some notion of how well-matched users and items are. We define the affinity score for user $\mathbf{u} = U^T_{i,:} \in \mathbb{R}^{d \times 1}$ and item $\mathbf{t} = T_{:, j} \in \mathbb{R}^{d \times 1}$ as:

\begin{equation}
    \text{Aff}(u, t) = \mathbf{u}^T\mathbf{t}.
\end{equation}

Prototype-based Matrix Factorization (ProtoMF) extends traditional matrix factorization by incorporating $L$ prototypes ($P_u \in \mathbb{R}^{L_u \times d}$ for users, and $P_t \in \mathbb{R}^{L_t \times d}$ for items) in potentially both embedding spaces of users and items. These prototypes act as anchor points in the latent factor space. Using these points, new embeddings ($U^* \in \mathbb{R}^{N \times L_u}$ and $T^* \in \mathbb{R}^{M \times L_t}$) are calculated for each user and item using the cosine similarity ($\text{sim}$) with respect to each prototype in $P_u$ and $P_t$.

The affinity between users and items can therefore be expressed using the new embedding coordinates $\mathbf{u}^*$ and $\mathbf{t}^*$:
\begin{equation}
    \text{Aff}(u, t) = {\mathbf{u}^*}^T \cdot \hat{t} + {\mathbf{t}^*}^T \cdot \hat{u}
    \label{eq:transformed_affinity}
\end{equation}

where $\hat{u}$ and $\hat{t}$ are linear transformations of $u^*$ and $t^*$ to create a common space where affinity scores can be calculated. The affinity score is then used as the final recommendation score of a user-item pair. The use of prototypes therefore provides extra degrees of freedom for representing users and items such to better calibrate affinity to the training examples.

\section{Methodology}
\begin{figure*}[th!]
    \centering
    \begin{subfigure}[b]{0.45\linewidth} 
        \centering
        \includegraphics[width=0.8\linewidth]{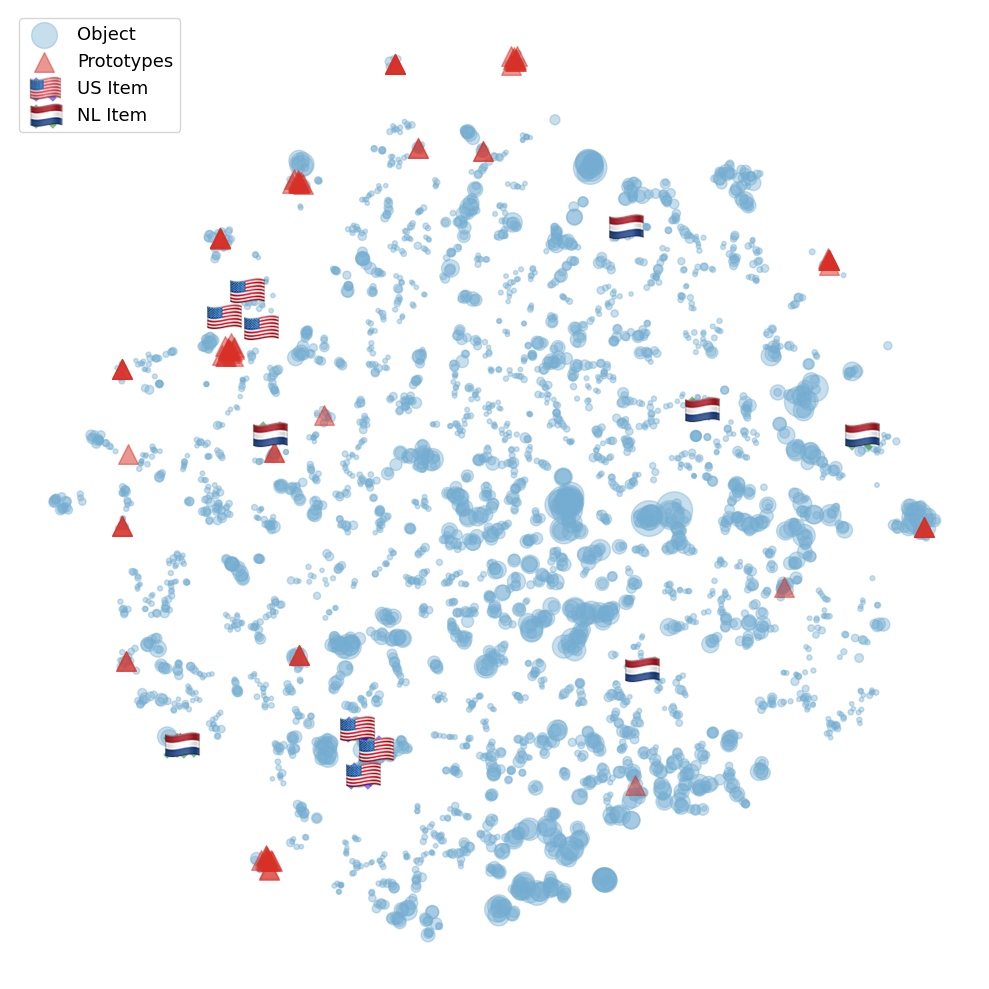}
        \caption{Prototypes (triangles) cluster more around popular items (US) than less popular items (NL).}
        \label{fig:item_interactions_b}
    \end{subfigure}
    \begin{subfigure}[b]{0.45\linewidth} 
        \includegraphics[width=0.8\linewidth]{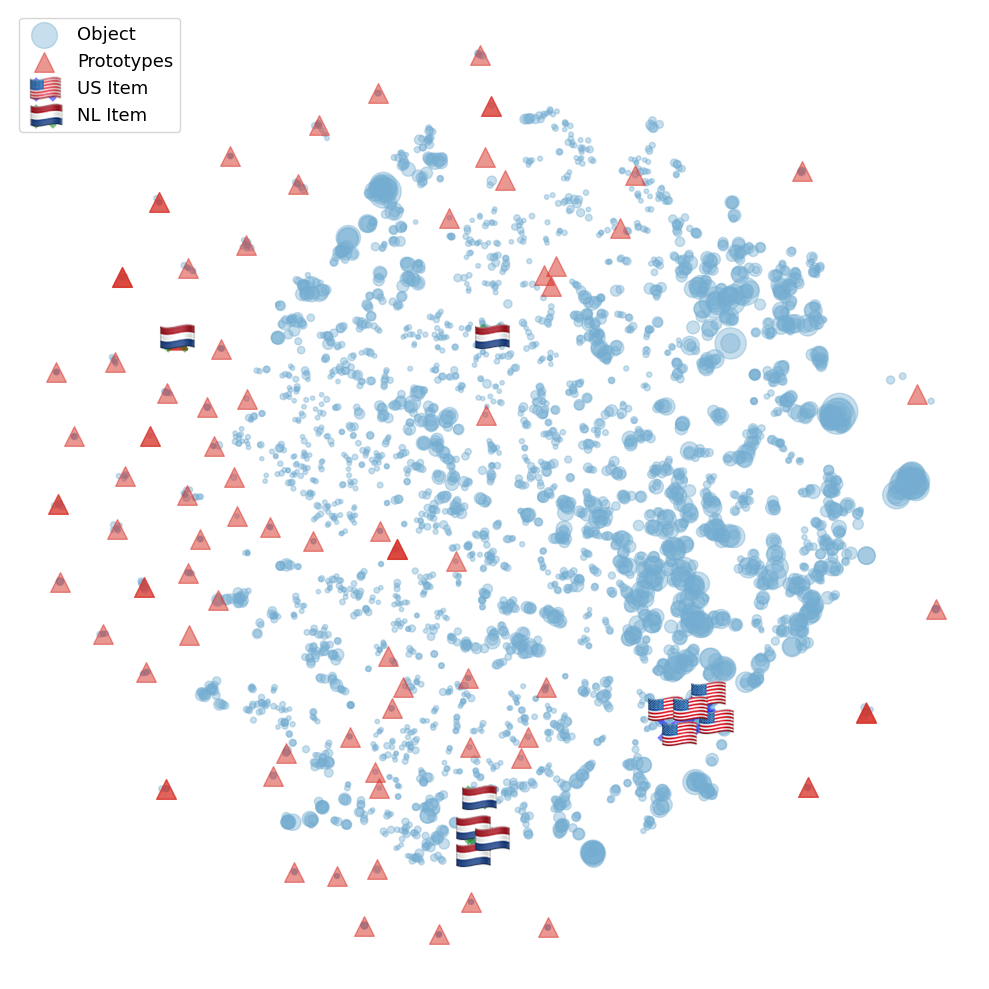}
        \caption{Prototypes (triangles) are more evenly distributed, and the less popular items (NL) become grouped.}
        \label{fig:item_interactions_c}
    \end{subfigure}
    \caption{t-SNE visualizations of the item embedding space for (a) the baseline ProtoMF model and (b) the model after applying both Prototype k-Filtering and the Prototype-Distributing Regularizer. Larger dots represent more popular items. The triangles indicate the positions of prototypes. NL and US represent an underrepresented and overrepresented country, respectively.}
    \label{fig:embeddings}
\end{figure*}

We introduce our novel contributions for enhancing the ProtoMF approach to mitigate both cultural and popularity bias. 
\subsection{Prototype K-filtering}
\begin{figure}[H]
  \centering
  \includegraphics[scale=0.4]{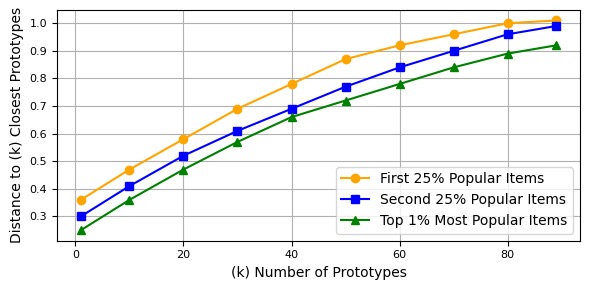} 
  \caption{Average distance of items to k nearest prototypes, grouped by popularity, for the baseline ProtoMF method. The gap in average distances between items of different popularity levels narrows with smaller k values.}
  \label{fig:pop_distances}
\end{figure}

Our first contribution, Prototype K-filtering, filters-out prototypes that are further from the item or user in their relative embedding spaces. As such, the model only uses the $k$ nearest prototypes for generating user and item representations during training, which is a hyperparameter that we heuristically train using \ref{fig:pop_distances} insights. This allows the prototypes to grasp more local information that relates to the closer neighboring examples in the embedding space, and forces the model to learn a less ``global'' set of prototypes. Formally, we can write the new embeddings as
\begin{equation}
u^{**} = \text{TopK} (u^*, k_u) \in \mathbb{\mathbb{R}}^{k_u}, k_u < L_u
\end{equation}
\begin{equation}
t^{**} = \text{TopK} (t^*, k_t) \in \mathbb{\mathbb{R}}^{k_t}, k_t < L_t
\label{eq:transformed_vectors}
\end{equation}

where the $\text{TopK}$ function outputs a $k$-dimensional feature vector for each user and item, which is representative of its $sim$ to the $k$ nearest prototypes.

\subsection{Prototype-Distributing Regularizer}
For our second approach, we add a regularization term to the loss function to spread the angular position of prototypes across the embedding space. Formally, we add the following to the original loss function:
\begin{equation}
      L = L_{original} + \lambda_u \cdot ||\hat{P}_{L_u} \hat{P}_{L_u}^T||_2^2 + \lambda_t \cdot ||\hat{P}_{L_t}  \hat{P}_{L_t}^T||_2^2
\end{equation}

Where $\hat{P}_{L_u} \in \mathbb{R}^{L_u \times d}$ and $\hat{P}{L_t} \in \mathbb{R}^{L_t \times d}$ are matrices representing the normalized prototype vectors for both user and item spaces, and $L_{original}$ is the original loss function. In these matrices, each of the prototype vectors has been normalized to have a L2-norm of 1. By enforcing orthonormality of the prototype vectors, we encourage the prototypes to serve as the basis vectors of the embedding space. This approach prevents prototypes from gravitating towards the same regions of the embedding space, thereby redistributing them to cover more marginal areas. This leads to more information gain, as it ensures a more effective and diverse utilization of the embedding space. This concept is akin to the discrete latent space representation in \citet{oord2018neural}'s work, where the effective use of embedding vectors is crucial for capturing the latent structure of the data.

\section{Empirical Evaluation}
In this section, we first outline our experimental set up, and then we answer two research questions to explore the effectiveness of our prototype-based contributions in mitigating cultural bias compare to existing approaches.

\paragraph{\textbf{Datasets}}
We conduct our experiments on two datasets. Firstly, we use a subset of the \emph{LastFM-2b dataset} \cite{MELCHIORRE2021102666} which consists of around 15,258 users, and 4,082,530 million songs with 30,357,786 number of interactions that we integrate into the learning process to avoid losing insights from the minority groups. In order to obtain the artists' country, we augment the dataset using \href{musicbrainz.com}{musicbrainz.com} data and matching the artists by name. Secondly, we also use the \emph{MovieLens-1M} \cite{harper2015movielens} dataset, which contains 1,000,209 ratings of 3,900 movies from 6,040 users and use the movie genre as the attribute to group the under and over-represented items.

\paragraph{\textbf{Baselines}}
To compare our model with existing work, we select four baselines: standard Matrix Factorization \cite{koren2009matrix}, anchor-based collaborative filtering \cite{barkan2021anchor}, and the standard prototype-based matrix factorization (ProtoMF) \cite{melchiorre2022protomf}. We include an additional benchmark to compare with existing popularity bias mitigation, ZeroSum, by incorporating \citet{rhee2022countering}'s method into ProtoMF to fairly compare its effectiveness with our own approach. We perform hyperparameter tuning to ensure optimal performance of each baseline and also our own contribution by tuning $K$ and $\lambda$.

For tuning our own models, we first identify the best-performing standard ProtoMF model through hyperparameter tuning. Subsequently, to maintain computational feasibility, we tune our models only for the new parameters introduced by our contributions, $k$ and $\lambda$, while keeping the other parameters fixed at their optimal values from the standard ProtoMF model.

\paragraph{\textbf{Fairness Metrics}}

We evaluate fairness of the output recommendations using the following metrics:
\begin{itemize}[leftmargin=*]

\item \textbf{Average Item Ranking}: Ranking is crucial for fairness as users interact differently based on item positions \cite{ursu2018power}. We report the average ranking of items in underrepresented ($\mu(r_{under})$) and majority groups ($\mu(r_{over})$) over model outputs of 100 items.
\begin{itemize}[leftmargin=*]
    \item In Last-FM, we define the majority group as the set of artists from the top three most popular countries by listen counts \{US, UK\}. The underrepresented groups are \{BR, NL, PL\}.
    \item For ML-1M, we categorize subgroups based on genres. The overrepresented genres are \{Drama, Comedy\}, while the underrepresented genres are \{War, Sci-fi, Western\}.
\end{itemize}

\item \textbf{Long-Tail Item Visibility}: We also report the proportion of long-tail items in the top 10 recommendations for all users, including repeated appearances. We define long-tail items as the top 10\% of items with the \textit{least} interactions, using the logarithmic interaction count discussed by  \citet{salganik2024fairness}.
\end{itemize}


\begin{table*}[ht!]
    \centering
    \caption{Performance and fairness results of baseline methods compared to prototype-$k$-filtering, Prototype-distributing regularization, and a combination of the two. Arrows indicate whether higher or lower values are more desirable.}
    \label{tab:experiments_baselines}

 \resizebox{\textwidth}{!}{ 
\begin{tabular}{|cc|cc|cc||cc|cc|cc|}
\hline

\multicolumn{2}{|c|}{\multirow{2}{*}{\textbf{Model}}}                             & \multicolumn{2}{c|}{\textbf{HitRatio@10} $\uparrow$}             & \multicolumn{2}{c||}{\textbf{NDCG@10} $\uparrow$}           & \multicolumn{2}{c|}{$\mathbf{\mu(r_{under})}$ $\downarrow$}            & \multicolumn{2}{c|}{$\mathbf{\mu(r_{over})}$}             & \multicolumn{2}{c|}{\textbf{LT Ranking} $\downarrow$}               \\ 

\cline{3-12} 
\multicolumn{2}{|c|}{}                                                   & \multicolumn{1}{c|}{LFM} & ML & \multicolumn{1}{c|}{LFM} & ML & \multicolumn{1}{c|}{LFM} & ML & \multicolumn{1}{c|}{LFM} & ML & \multicolumn{1}{c|}{LFM} & ML  \\ 

\hline

\multicolumn{2}{|c|}{MF \cite{koren2009matrix}}                                                 & \multicolumn{1}{c|}{0.159}      & 0.452     & \multicolumn{1}{c|}{0.083}      & 0.248     & \multicolumn{1}{c|}{49.299}      & 56.041     & \multicolumn{1}{c|}{49.755}      & 49.226     & \multicolumn{1}{c|}{25.991}      & 35.842    \\ 
\hline

\multicolumn{2}{|c|}{ACF \cite{barkan2021anchor}}                                              & \multicolumn{1}{c|}{0.569}      & 0.618     & \multicolumn{1}{c|}{0.317}      & 0.345     & \multicolumn{1}{c|}{53.970}      & 57.198     & \multicolumn{1}{c|}{43.458}      & 49.652     & \multicolumn{1}{c|}{20.009}      & 18.142      \\ 
\hline

\multicolumn{2}{|c|}{ProtoMF \cite{melchiorre2022protomf}}                                           & \multicolumn{1}{c|}{0.581}      & 0.656     & \multicolumn{1}{c|}{0.337}      & 0.381     & \multicolumn{1}{c|}{51.106}      & 54.061     & \multicolumn{1}{c|}{44.300}      & 49.510    & \multicolumn{1}{c|}{29.008}      & 36.365   \\ 
\hline

\multicolumn{2}{|c|}{ZeroSum \cite{rhee2022countering}}                                           & \multicolumn{1}{c|}{0.564}      & 0.615     & \multicolumn{1}{c|}{0.327}      & 0.347     & \multicolumn{1}{c|}{52.233}      & 53.513     & \multicolumn{1}{c|}{44.733}      & 49.339     & \multicolumn{1}{c|}{21.543}      & 23.669     \\ 
\hline 

\multicolumn{1}{|c|}{\multirow{3}{*}{Our Models}} & \textbf{Prototype $k$-Filtering ($k_t$)}    & \multicolumn{1}{c|}{0.594}      & 0.651    & \multicolumn{1}{c|}{0.344}      &  0.379    & \multicolumn{1}{c|}{49.849}      & 53.984     & \multicolumn{1}{c|}{43.852}      & 49.780    & \multicolumn{1}{c|}{ 21.995}      & \textbf{16.375}      \\ 
\cline{2-12} 
 \multicolumn{1}{|c|}{}                           & \textbf{Prototype-Distributing Regularizer} ($\lambda_t$) & \multicolumn{1}{c|}{0.598}      & 0.660      & \multicolumn{1}{c|}{0.348}      & \textbf{0.386}     & \multicolumn{1}{c|}{51.111}      & 53.486     & \multicolumn{1}{c|}{44.085}      & 48.774     & \multicolumn{1}{c|}{18.995}      & 18.369        \\
\cline{2-12} 
\multicolumn{1}{|c|}{}                           & \textbf{$k$-Filtering and Distributing ($k_t$, $\lambda_t$)}                & \multicolumn{1}{c|}{\textbf{0.600}}      & \textbf{0.657}     & \multicolumn{1}{c|}{\textbf{0.352}}      & 0.383     & \multicolumn{1}{c|}{\textbf{49.552}}      & \textbf{51.600}     & \multicolumn{1}{c|}{44.296}      & 49.678     & \multicolumn{1}{c|}{\textbf{18.013}}      & 18.135      \\ 
\hline

\end{tabular}}
\end{table*}

\paragraph{\textbf{RQ1: Do prototypes in the embedding space of ProtoMF possess inherent properties that can lead to cultural bias?}}

To answer this question, we analyze the embedding space of the baseline ProtoMF method. 
The baseline embedding space in Figure~\ref{fig:pop_distances} indicates that popular items tend to be closer to prototypes. Proximity to prototypes provides an advantage to items, since it increases the cosine similarity values in its representation (Equation \ref{eq:transformed_vectors}), and therefore the chance of a higher affinity score with users (Equation \ref{eq:transformed_affinity}). This observation provides an initial justification for using a subset of prototypes for recommendations to enhance fairness.  

Not only do prototypes in the baseline model optimize for proximity to popular items, but they also concentrate together in these regions (see Figure \ref{fig:item_interactions_b}). This concentration exacerbates the issue of providing advantages to more popular items, leading to further disparities in rankings between popular and less popular items. This could leave many regions in the embedding space without representative prototypes, leading to systemic unfairness for less popular items, and therefore cultures. \\



\paragraph{\textbf{RQ2: How successful are our proposed approaches, which modify the learning process of the embedding space to enhance the discoverability of minority cultures and mitigate popularity bias, compared to existing methods?}}

To answer this question, we analyze the impact of our contributions on the embedding space. 

\begin{figure}[htbp]
    \centering
    \includegraphics[width=\linewidth]{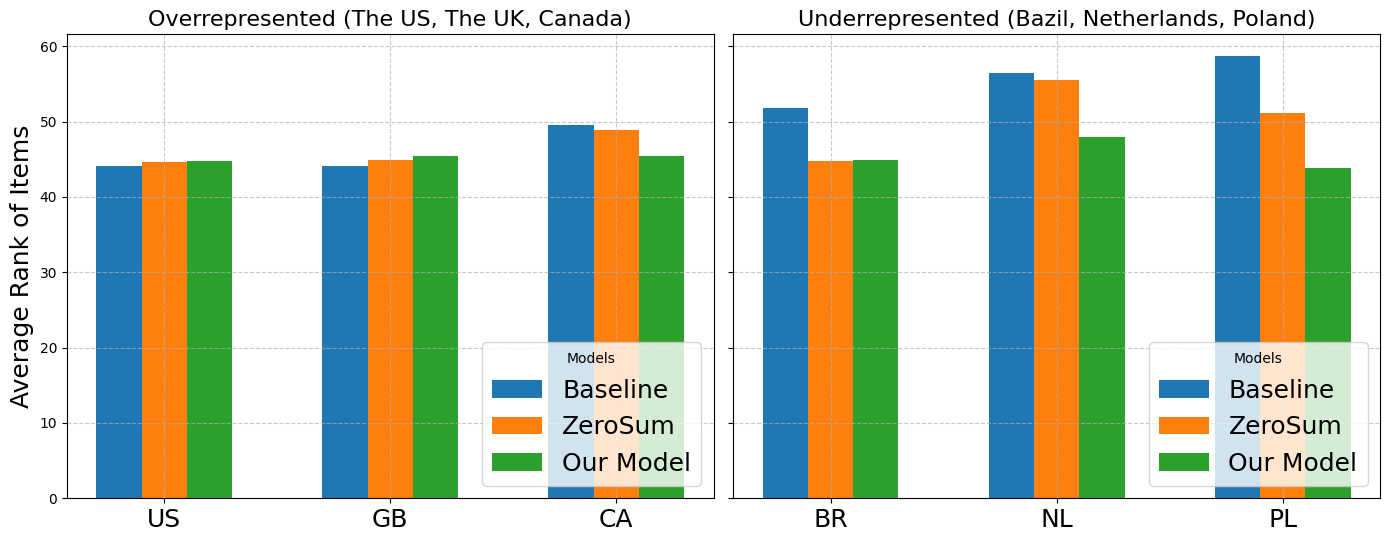}
    \caption{Average rank of overrepresented items (from US, GB, CA) and underrepresented items (from BR, NL, PL), comparing ProtoMF, ZeroSum and our model using $k$ and $\lambda$.}
    \label{fig:rank_vs_popularity}
\end{figure}

Figure~\ref{fig:embeddings} illustrates the difference in prototype placement in the embedding space between the baseline ProtoMF model (\ref{fig:item_interactions_b}) and our model with the regularizer (\ref{fig:item_interactions_c}). The regularization increases the space between prototypes, reducing the dominance of popular items in the embedding space.
Moreover, the effect of our model on popularity bias can be seen in the Figure \ref{fig:rank_vs_popularity}. Our contributions forces less popular items to have a lower ranking (higher on the recommendation list), and popular items to appear in higher ranks (lower on the recommendation list). Our model therefore provides greater visibility to less popular items than the baseline model \cite{ursu2018power}.

The quantitative results of the contributions, comparing variations of our best-performing models against existing baselines, can be found in Table \ref{tab:experiments_baselines}. We report recommendation performance metrics, including HitRatio@10 and NDCG@10, as well as fairness metrics. Our findings reveal that our models outperform standard MF in terms of performance and are on par with ProtoMF and ACF on both datasets. Additionally, our models successfully reduce the difference between the average rank for under- and over-represented groups, demonstrating improved fairness.  Furthermore, they are more inclusive towards less popular items, as evidenced by the decrease in the average long-tail rank. Notably, while both k-filtering and the regularizer independently improve fairness, their combined effect amplifies these improvements, leading to the fairest outcomes.

\section{Conclusion \& Future Directions}

In this work, we introduced two novel enhancements to the ProtoMF model's embedding space, enabling us to effectively mitigate cultural and popularity bias in music recommendation. Crucially, these enhancements operate independently of the model's original loss function. This decoupling of fairness and accuracy optimization is a key advantage of our approach, as it allows us to directly target and mitigate biases without the inherent trade-offs that often plague methods that modify the loss function. Our results demonstrate that this approach is not only effective in significantly improving fairness metrics, but also maintains, and in some cases even enhances, the overall recommendation quality across both datasets.

Further extensions to this work include expanding the evaluation across diverse music datasets with broader cultural and genre representation to enhance our understanding of the model's generalizability, especially given the limited data available on non-Western countries. Future work will focus on cultural matchings of users and items, considering the behaviors of various cultures as highlighted by \citet{lesota2022traces}, and emphasize user-side fairness. Additionally, we plan to investigate prototype collisions in the embedding space using explainability techniques to understand and theorize these collisions beyond popularity bias measures, which will also apply to the new embedding space after training our updated models. Lastly, we also acknowledge the ethical considerations surrounding the use of cultural attributes in recommender systems, as discussed in Appendix \ref{sec:ethical-considerations}.

\section*{Acknowledgments}
Funding support for project activities has been partially provided by Canada CIFAR AI Chair, and Facebook award. We also express our gratitude to Compute Canada for their support in providing facilities for our evaluations.

\bibliographystyle{ACM-Reference-Format}
\bibliography{main}

\newpage

\appendix

\section{Ethical Considerations and Broader Impacts}
\label{sec:ethical-considerations}

While this research did not involve direct interaction with human subjects or the collection of new data, it utilizes existing datasets (LastFM-2b and MovieLens-1M) that represent the behaviors and preferences of real users. We acknowledge that these datasets, while anonymized, may contain inherent biases or reflect societal inequalities. We strive to use these datasets responsibly and critically analyze any potential biases in our findings.

Our work aims to mitigate cultural bias in music recommender systems, which has important implications for fairness and equity in the digital music landscape. By promoting greater visibility for artists from underrepresented cultural groups, our methods could potentially:
\begin{itemize}
    \item Empower creators from marginalized communities
    \item Enhance cultural diversity in music consumption
    \item Counteract existing biases in the music industry
\end{itemize}

However, we also acknowledge potential ethical concerns:

\begin{itemize}
    \item Oversimplifying culture: In our analysis, we used country of origin as a proxy for cultural background. While this provides a useful starting point, it's important to acknowledge that culture is a complex and multifaceted concept that cannot be fully captured by a single attribute like nationality. This simplification could potentially lead to inaccurate assumptions or generalizations about the cultural preferences.
    \item Unintended consequences: Our methods could inadvertently introduce new biases or reinforce existing ones if not carefully evaluated and monitored.
    \item Overemphasis on culture: Focusing solely on cultural fairness could potentially neglect other important factors like individual preferences or musical quality.
    \item Limited generalizability: Our findings may not generalize to other recommender systems or cultural contexts.
\end{itemize}
To address these concerns, we emphasize the importance of ongoing research and continuous evaluation of fairness metrics in deployed recommender systems. We also advocate for transparency in the design and implementation of these systems, allowing users to understand and control how recommendations are generated.

\newpage
\section{Complimentary Experiment Results}
In Appendix B (Table \ref{tab:all_experiment_results}), we present a comprehensive set of results across various combinations of hyperparameters $k$ (prototype filtering) and $\lambda$ (regularization strength) for the ProtoMF model on the LastFM dataset. The table includes performance metrics (HitRatio@10, NDCG@10). These detailed results allow for a deeper exploration of the impact of different hyperparameter values on recommendation quality.

\begin{table}[h]
    \centering
    \caption{Performance and Fairness of ProtoMF on the LastFM-2b Dataset Across Different Combinations of $k$ (Prototype Filtering) and $\lambda$ (Regularization Strength), where  $k=-1$ indicates using all prototypes.}
    \label{tab:all_experiment_results}
    \begin{tabular}{c|c|c|c|c|c}
        \toprule
        \( K_u \) & \( \lambda_u \) & \( K_t \) & \( \lambda_t \) & HitRatio@10 & NDCG@10  \\
        \midrule
        
        -1 & 0.0 & 60 & 0.0 & 0.593 & 0.348\\ 
        \hline
                
        60 & 0.003 & 60 & 0.1 & 0.577 & 0.338 \\ 
        \hline
                
        60 & 0.0 & 25 & 0.003 & 0.569 & 0.332 \\ 
        \hline
                
        -1 & 0.0 & 25 & 0.003 & 0.569 & 0.329 \\ 
        \hline
                
        -1 & 0.0 & 60 & 0.9 & 0.562 & 0.326 \\ 
        \hline
                
        -1 & 0.0 & 40 & 0.0 & 0.559 & 0.321 \\ 
        \hline
                
        12 & 1.0 & 12 & 0.1 & 0.552 & 0.323 \\ 
        \hline
                
        -1 & 0.0 & 60 & 0.25 & 0.552 & 0.320 \\ 
        \hline
                
        -1 & 0.0 & 12 & 0.0 & 0.543 & 0.310 \\ 
        \hline

        -1 & 0.0 & 60 & 0.387 & 0.534 & 0.307 \\ 
        \hline

        -1 & 0.1 & 25 & 0.0 & 0.519 & 0.301 \\ 
        \hline

        60 & 0.003 & 60 & 0.003 & 0.515 & 0.301 \\ 
        \hline

        25 & 0.0 & 25 & 10.0 & 0.513 & 0.294 \\ 
        \hline

        -1 & 10.0 & 40 & 40 & 0.505 & 0.291 \\ 
        \hline

        -1 & 0.0 & 12 & 0.0 & 0.488 & 0.273 \\ 
        \hline

        \bottomrule
    \end{tabular}
\end{table} 
\end{document}